\documentclass[pra,aps,hyperref,twocolumn,showpacs]{revtex4}

\usepackage{graphicx}
\usepackage{dcolumn}
\usepackage{bm}

\newcommand{\rtw}{\longrightarrow}
\def\veps{\varepsilon}

\begin{document}

\title{Proposed search for T-odd, P-even interactions in spectra of chaotic atoms}

\author{Muir J. Morrison}
\affiliation{Department of Physics, University of Nevada, Reno, Nevada 89557,
USA}

\author{Andrei Derevianko}
\affiliation{Department of Physics, University of Nevada, Reno, Nevada 89557,
USA}

\begin{abstract}
Violation of fundamental symmetries in atoms is the subject of intense experimental and theoretical interest. P-odd, T-even transitions have been observed and are in excellent agreement with electroweak theory. Searches for permanent electric dipole moments have placed bounds on T-odd, P-odd interactions, constraining proposed extensions to the Standard Model of elementary particles. Here we propose a new search for T-odd, P-even (TOPE) interactions in atoms. We consider open-shell atoms, such as the rare-earth atoms, which have dense, chaotic excitation spectra with strong level repulsion. The strength of the level repulsion depends on the underlying symmetries of the atomic Hamiltonian. TOPE interactions lead to enhanced  level repulsion. We  demonstrate how a statistical analysis of many chaotic spectra can determine the strength of level repulsion; in particular, the variance of the number of levels in an energy range has been shown to be a useful measure. We estimate that, using frequency comb spectroscopy, a sufficient number of chaotic levels could be measured to match or exceed the current experimental bounds on TOPE interactions.
\end{abstract}


\maketitle

\section{Background}
\label{Sec:Background}

Electromagnetic interactions are invariant under any combination of the
discrete symmetry operations, including spatial parity (P), time-reversal (T), and
charge conjugation (C). Conservation of these symmetries lead to the well-known selection rules for atomic
expectation values and transition amplitudes. Searching for violation of these selection
rules enables revealing minute non-electromagnetic corrections to the atomic
Hamiltonian. First observation of the parity-forbidden E1 transition in
bismuth was an important confirmation of the electro-weak theory in
semi-leptonic sector~\cite{BarZol78}. Later atomic experiments with cesium
confirmed low-energy limit of the electro-weak theory at the level of
radiative corrections~\cite{WooBenCho97,PorBelDer09,Porsev2010}. 
While the parity violation is firmly established, the T-violation is probed  in
searches for the electric dipole moments (EDMs) of elementary
particles, which violate both P and T symmetries~\cite{KhrLam97}. The most
stringent limits to date on the electron EDM were derived from experiments with
atomic thallium \cite{RegComSch02} and with YbF molecule \cite{HKSS11}. 
The most accurate limit on atomic EDM comes from Hg experiment~\cite{Griffith2009}. It is worth emphasizing, that while
no EDM has been discovered yet, 
the current limits on eEDM are very powerful. These limits have 
ruled out many extensions to the Standard model
including minimal supersymmetry and severely constrained SUSY models in
general.

Here we focus on the T-odd, P-even (TOPE) interactions. 
The TOPE interactions are far less explored than the described  P-odd and P,T-odd
couplings. 
One reason for this paucity is of theoretical nature: phenomenological Lagrangians for TOPE
interactions involve field derivatives; thereby in most popular models
such terms appear only as radiative corrections. Another reason is purely
experimental: there is no convenient observable, like EDM, associated with
TOPE interactions. This makes designing experiments more challenging. So far only two
types of atomic physics experiments were suggested and carried out  for TOPE interaction searches.
Here we propose an alternative to these searches.

Before describing our proposal, we review the previous efforts and the established limits.
The positronium experiment~\cite{Conti1993} relied on the compelling arguments of the CPT theorem which implies 
that non-vanishing TOPE interactions would violate charge conjugation symmetry.
The experiment probed the C-forbidden single-photon
transition between the $2{}^3\!S_1$ and $2{}^1\!P_1$ states of positronium. That
lead to a constraint on the electron-positron TOPE
interaction $H^\mathrm{e-\bar{e}}_\mathrm{T}$:
 \begin{equation}\label{Ps1}
 \frac{\langle 2{}^1\!P_1| H^\mathrm{e-\bar{e}}_\mathrm{T}|2{}^3\!P_1\rangle}
 {E_{{}^1\!P_1}-E_{{}^3\!P_1}}
 =0\pm 0.036\,.
 \end{equation}

The second search~\cite{HoBa02}, proposed by \citet{Kozlov1989},  focused on the T-odd
correlation $\bm{k}\cdot\bm{E}$ in the refractive index of atomic vapour near
the $6p_{1/2} \rtw 6p_{3/2}$ transition in thallium. Here $\bm{k}$
is photon momentum and $\bm{E}$ is external DC electric field. 
The experiment on Tl was carried out on a particular hyperfine transition and
placed a limit on the electron-proton interaction (with valence proton
in Tl nucleus).  One  can think of this interaction as a T-odd hyperfine structure.

The experiment~\cite{HoBa02} placed the following limit on the nuclear-spin-dependent (NSD) interaction:
 \begin{equation}\label{Tl1}
 \frac{\langle 6p_{1/2}| H^\mathrm{NSD}_\mathrm{T}|6p_{1/2}\rangle}
 {E_{6p_{3/2}}-E_{6p_{1/2}}}
 =(0.9\pm 2.0)\times 10^{-3}.
 \end{equation}

Up to now (\ref{Ps1}) and (\ref{Tl1}) are the only {\em direct} experimental limits on TOPE interaction in
atoms. Stronger {\em indirect} model-dependent limits can be obtained from the EDM experiments.
Indeed, in the second order of  perturbation theory the T-odd and P-odd
interactions generate effective P,T-odd interaction. This, in turn, contributes to
the EDMs of elementary particles and atoms. Khriplovich noted that for the
short-range interaction this mechanism leads to very
stringent bounds on the short-range TOPE interaction \cite{Khr91,CK92}. For
the long-range TOPE interaction such indirect limits are less stringent, but
still stronger, than the direct limits (\ref{Ps1},\ref{Tl1}) \cite{HoBa02}. Note,
however, that the indirect limits are model-dependent, i.e., they depend on assumptions with respect to employed phenomenological Lagrangians.


{
}

In nuclear physics, the constraint on TOPE comes from level-repulsion statistics in chaotic nuclear spectra~\cite{FrKoPa85,FrKoPa87}. For such a system, the Hamiltonian may be represented as a block-diagonal matrix, where each block corresponds to states of the same total angular momentum and parity. If we are merely interested in statistics of level spacings, we can avoid diagonalizing or even constructing the Hamiltonian. Instead one could consider ensembles of random matrices whose matrix elements are independent random variables (for nuclear physics applications, see~\cite{Weiden2009}). These random variables are described by Gaussian distribution with a fixed variance. Ensemble averages of fluctuation measures (such as the $\Sigma^2$ measure used in Ref.~\cite{FrKoPa85}) agree well with experimental data, even despite unrealistic many-body interactions such matrices contain.

More generally, any open-shell system of interacting fermions exhibits quantum chaos. Technically, any quantum system with a nonseparable wave equation is chaotic; this includes even such simple systems as neutral helium~\cite{Haake2010}. For our purposes, this is much too broad a criteria. We are interested in atoms with chaotic spectra that are also dense (containing a statistically large number of levels). We also require strong interactions between valence electrons so that total angular momentum and parity are the only ``good" quantum numbers (see Sec.~\ref{Sec:sensAndExpt} as well as~\cite{Rosenzweig1960} for further discussion of this point).

For these reasons, the rare-earth atoms are the ideal candidates for our search. Their Hamiltonians can be modeled with random matrix ensembles exactly as described above for nuclei. Practically, the most useful criteria to distinguish between chaotic and regular (integrable) quantum systems is the distribution of nearest neighbor spacings of energy levels. Regular systems exhibit an exponential distribution for nearest-neighbor spacings, while chaotic systems follow the Wigner distribution~\cite{Haake2010,Rosenzweig1960}.

The basic idea of this paper is to point out the utility of the level-repulsion statistics method of nuclear physics for searching for TOPE interactions in atomic physics. 


\section{Method}
Spectra in nuclei and complex open-shell atoms exhibit quantum chaos, characterized 
as described in Sec.~\ref{Sec:Background}. We can model such many-particle systems with random matrix ensembles (RMEs). If T-reversal is preserved, all the Hamiltonian matrix elements may be taken as {\em real} random variables, whereas if T-reversal is broken, the matrix elements are in general {\em complex}. The real and imaginary parts of $U_{ij}$  are independent random variables. Therefore, as described in the appendix,  T-violating interactions lead to enhanced level repulsion (quadratic, rather than linear, for small spacings).

To search for TOPE interactions, following~\cite{FrKoPa85}, we use a RME of the form 
\begin{equation}
\{H_\alpha\} = \{H(S) + i\alpha H(A)\} \,. 
\end{equation}
Here the $\{H(S)\}$ matrix  is real and symmetric, representing the dominant electromagnetic interactions, 
and $\{H(A)\}$ is real and anti-symmetric, characterizing TOPE interactions.  We assume that 
both random matrices have the same variance $v^2$ for all matrix elements, with the real parameter $\alpha \ll 1$ determining the relative strength of TOPE and electromagnetic interactions. The $\alpha=0$ case is conventionally termed the Gaussian orthogonal ensemble (GOE), while $\alpha=1$ is known as the Gaussian unitary ensemble (GUE), in reference to the symmetry groups their respective Hamiltonians posess.
From perturbation theory, it should appear plausible that $\Lambda$, defined by 
\begin{equation}
\Lambda = \alpha^2v^2/D^2(E)
\end{equation}
is a more useful parametrization of the relative strength of the TOPE interactions than $\alpha$; one may think of it as the square of the expected value of the perturbation (although technically, $\langle H(A) \rangle = 0$ since $H(A)$ is antisymmetric).

Before we can compare experimental spectra with RME predictions, we must ``unfold" the spectra. This is necessary because RMEs have a uniform level density (alternately, the average level spacing is uniform) whereas real spectra clearly do not. To unfold the spectra, one would first construct the number staircase function $\mathcal{N}$. Formally, it is given by~\cite{Weiden2009}
\begin{equation}
\mathcal{N}(E)=\int_{-\infty}^E \sum\limits_{i} \delta(E^\prime-E_i)\, dE^\prime,
\label{Eq: N}
\end{equation}
i.e., $\mathcal{N}(E)$ is simply the number of levels with energy less than or equal to $E$. By fitting a polynomial of reasonable order to $\mathcal{N}(E)$, one would obtain $\mathcal{N}_{fit}(E)$, and differentiating this with respect to $E$ would give the average level density $\rho(E)$~\footnote{Note that the average level spacing $D(E)$ is simply the inverse of $\rho(E)$.}. The final step would be to map each old eigenvalue $E_i$ to a new eigenvalue $\varepsilon_i$ with the formula $\varepsilon_i = \int_{-\infty} ^{E_i} \rho(E) dE$. But since $\rho(E) = \frac{d}{dE}\mathcal{N}_{fit}(E)$, the unfolded eigenenergies $\varepsilon_i$ are simply given by
\begin{equation}
\varepsilon_i = \mathcal{N}_{fit}(E_i),
\label{Eq: N}
\end{equation}
so that the unfolded energies are dimensionless with uniform average level spacing equal to unity.

We now introduce statistical measures with which we may compare unfolded data to the predictions of the RMEs. For our purposes, the most useful statistic is $\Sigma^2(r)$, the variance of the number of levels in an energy interval that contains $r$ levels on average. (Note that the unfolded spectra is dimensionless with average level spacing equal to unity, so an energy interval of length $r$ contains $r$ levels on average.) Define $n(\varepsilon,r)$ 
to be the number of levels in a small energy interval of length $r$ at energy $\varepsilon$ (i.e., the number of levels between $\varepsilon$ and $\varepsilon +r$). Then $\Sigma^2(r)$ is given by
\begin{equation}
\Sigma^2(r) = \overline{n^2(\varepsilon,r)}-{\overline{n(\varepsilon,r)}}^2 = \overline{n^2(\varepsilon,r)}-r^2,
\end{equation}
where the overbar denotes a running average over the measured spectra; note $\overline{n(\varepsilon,r)} = r$ only for unfolded data.

In most nuclear physics applications, a different statistic known as the spectral rigidity, $\Delta_3$, is preferred over $\Sigma^2(r)$, primarily because $\Delta_3$ is smoother and fits data to a GOE more easily~\cite{Weiden2009}. Qualitatively, $\Delta_3$ measures a spectrum's deviation from the best-fit uniform spectrum. The two statistics are related however, and in fact $\Delta_3$ can be expressed as an integral over $\Sigma^2(r)$~\cite{Haq1982}. However, $\Delta_3$ also ``washes out" the distinction between GOE and GUE at small spacings where TOPE would be most evident, so for our purposes $\Sigma^2(r)$ is more useful.

The crucial point is that TOPE interactions make $\Sigma^2(r)$ smaller because of greater level repulsion. Lengthy derivations~\cite{French1988} demonstrate that for the relevant case of small $\Lambda$, 
\begin{equation}
\Sigma^2(r,\Lambda) \approx \Sigma^2(r,0)-4\Lambda
\end{equation}
i.e., the variance decreases and spectral uniformity increases as level repulsion is increased from the T-preserving case $\Sigma^2(r,0)$.

{


}

\section{Sensitivity and experimental realization}
\label{Sec:sensAndExpt}

We now consider how stringent a bound such a test could place on $\alpha$. The sample error $\sigma$ is simply a function of the number of levels considered $p$, so from the central limit theorem we would expect to have $\sigma \propto {p}^{-1/2}$. Detailed calculations verify this~\cite{French1988}; in fact, $\Sigma^2(1)$ has a $\chi^2_p$ distribution, so the sample error is simply $\sigma \approx0.6(2/p)^{1/2}\Sigma^2(1,\Lambda)$~\footnote{The factor of $0.6$ reduction in $\sigma$ comes from using partially overlapping intervals, as suggested in~\cite{French1988}.}. Roughly, $\Sigma^2(1,\Lambda) \approx 1/2$~\cite{French1988}. Therefore, to obtain a bound on $\alpha$ at the $10^{-3}$ level, comparable to the experimental bound from Eq.~(\ref{Tl1}) and an order of magnitude more stringent than the positronium bound from Eq.~(\ref{Ps1}), would require $\sim 10^5$ levels.

How many chaotic atomic states can we expect to find? Obviously there are infinitely many levels of a given $J$ and parity including Rydberg series, but we are only interested in chaotically mixed compound states. The authors of Ref.~\cite{DzuFla10} performed configuration interaction calculations on Th in order to estimate density of states. For Th II, they estimate that there are $\sim 10^3$ compound states with $J=3/2$ and even parity. Except for a few low-lying states, nearly all of these levels are chaotic and suitable for our purposes. We assume a comparable number of levels would be found for different angular symmetries and in different atoms. Spectra could be measured for the neutral and singly ionized lanthanides (perhaps even doubly-ionized), giving perhaps $10-30$ species depending on experimental challenges. If $\sim 10$ sequences with different angular symmetry are measured in each species, achieving $\gtrsim 10^5$ levels seems optimistic but plausible. 


Accurately locating $\sim10^5$ levels presents a daunting task, but it should be possible with reasonable efficiency using frequency comb (FC) spectroscopy. Prior works have used the molecular fingerprinting technique to measure molecular spectra in a massively parallel fashion (see, e.g.,~\cite{Thorpe2006, Diddams2007, Gohle2007}, or~\cite{Maddaloni2009, Foltynowicz2011} for surveys of experimental methods). Essentially, these experiments use the many teeth of the FC like thousands of cw-lasers simultaneously probing the sample. The same technique could be applied to the dense level structure of the lanthanides. However, for our purposes we need spectra with no spurious or missing levels, which demands FC coverage from the 
visible to the mid-UV, and perhaps even into the vacuum-UV or extreme-UV for ionized species. Such FCs already exist in the visible and near-IR, and rapid progress has been made in extending frequency combs to the mid-IR~\cite{Adler2010}, to the VUV~\cite{Yost2009}, and even into the XUV~\cite{Cingoz2012, Lee2011, Pinkert2011}. As this technology continues to improve, measuring the necessary spectra seems quite achievable.

Even with FC spectroscopy, making accurate angular momentum and parity assignments for so many levels would be challenging. It has been suggested~\cite{French1988} that a Bayesian analysis could help in this task, since the approximate form for the spacing distribution is known. This would introduce a few spurious and missing levels. Depending on the experimental challenges, however, it may be useful for labeling $10^5$ levels. Studies in nuclear spectra have shown that the effect of spurious and missing levels can be accounted for, and as long as their number is sufficiently small, they do not invalidate the predictions of RME theory~\cite{Weiden2009, French1988}.

{
}
Finally, we note that this proposal is best suited to a search for 
nuclear-spin independent TOPE interactions, experimentally and theoretically. Experimentally, the bound on 
nuclear-spin dependent TOPE from Eq.~(\ref{Tl1}) is currently an order of magnitude stronger than that on $e$-$\overline{e}$ TOPE interactions from Eq.~(\ref{Ps1}). The theoretical reasons are as follows. When we consider the Hamiltonian to be block-diagonal, there is some ambiguity as to what we consider to be the total angular momentum. First, let us neglect the nuclear spin and the hyperfine interaction. In the lanthanides, total orbital and spin angular momenta $L$ and $S$ are not ``good" quantum numbers thanks to strong spin-orbit coupling. In fact, the total electronic angular momentum $\vec{J}=\vec{L}+\vec{S}$ is the only good quantum number, along with parity, so we may assign blocks according to a particular $J$ and parity. Nearest-neighbor-spacings within a block will follow the Wigner distribution, indicating chaos~\cite{Weiden2009, Rosenzweig1960}. One may then compute fluctuation measures (i.e., $\Sigma^2(r)$) in each block, searching for deviations from GOE statistics. Any detected TOPE interaction would clearly be nuclear-spin independent. 

Alternately, if we include hyperfine interactions, the true total angular momentum of the system is $\vec{F}=\vec{J}+\vec{I}$, where $I$ is the nuclear spin. Then we may choose blocks according to a particular $F$ and parity. 
However, $J$ is still very nearly a good quantum number, which qualitatively means that states of a particular $F$ are not chaotically mixed. As shown in Ref.~\cite{Rosenzweig1960}, this means that the nearest neighbor distribution for a particular $F$ block will not be the Wigner distribution, but rather a superposition of several Wigner distributions corresponding to the different possible $J$ values. This distribution will interpolate between a Wigner and exponential distribution. Since the level statistics would not correspond to the GOE even approximately, it would be impossible to search for a TOPE interaction represented by a small admixture of GUE. While the level positions would certainly be affected by the existence of some nuclear-spin-dependent TOPE interaction, such an interaction would not be detectable from level statistics using the methods of this paper.


We would like to thank V. Dzuba, V. Flambaum, G. Gribakin, M.G. Kozlov, J. Lawler, K. A. Mitchell, and M. Pospelov for discussions. This work was supported in part by the NSF.

\appendix*
\section{}
\label{app}

In this appendix we give an illustrative derivation of the behavior
of the level-spacing distribution for the T-even and T-odd Hamiltonians for small level separations.
 As long as we are interested in very small energy splittings $\Delta$  between neighboring levels, 
we can neglect all other states and consider just these two levels. The random-matrix Hamiltonian then reads,
 \begin{equation}\label{a1}
  H = \left(
  \begin{array}{cc}
   a/2 & b+ic\\
   b-ic & -a/2
   \end{array}\right)\,.
  \end{equation}
For the T-even Hamiltonian the parameter $c$ can be set to zero, while for the
T-odd Hamiltonian all three parameters are nonzero. By diagonalizing this matrix we find the 
splitting to be 
$$
   \Delta = 2\sqrt{a^2+b^2+c^2}\,.
$$

Let us assume that all nonzero parameters are random variables with zero
average and the same variance $\sigma^2$. Then the probability that $\Delta < \veps \ll
\sigma$ is $\sim (\veps/\sigma)^3$ for T-odd Hamiltonian and  $\sim
(\veps/\sigma)^2$ for T-even Hamiltonian. The probability density
$\rho(\veps)$ is then $\sim (\veps/\sigma)^2$ and $\sim (\veps/\sigma)$
respectively. Clearly, the level statistics is affected by underlying symmetry of the Hamiltonian.


\end{document}